\documentclass[twocolumn,10pt]{article}
\pdfoutput=1

\usepackage[letterpaper,margin=2.25cm]{geometry}
\usepackage[T1]{fontenc}
\usepackage{amsmath,amssymb,amsfonts}
\usepackage{amsthm}

\usepackage{newtxtext,newtxmath}
\usepackage{graphicx}
\usepackage{booktabs}
\usepackage{multirow}
\usepackage{makecell}
\usepackage{xcolor}
\usepackage{bm}
\usepackage{float}
\usepackage{enumitem}
\usepackage{cite}
\usepackage{caption}
\usepackage{url}
\usepackage[colorlinks=false,pdfborder={0 0 0}]{hyperref}
\usepackage{balance}

\newtheorem{remark}{Remark}

\begin{document}

\twocolumn[{%
\begin{center}
{\large\bfseries Drift-Aware Multi-Target Space Tug Logistics Using Natural Orbital Precession\par}
\vskip 1.5em
{\normalsize
\textbf{Omer Burak Iskender}\textsuperscript{1,*},\
\textbf{Leonard Felicetti}\textsuperscript{2},\
\textbf{Kaan Gokturk}\textsuperscript{3},\
\textbf{David Tellett}\textsuperscript{4},\
\textbf{Adam Baker}\textsuperscript{4}\par}
\vskip 1.5em
{\normalsize\textsuperscript{1}\,\textit{Nanyang Technological University, Singapore.} E-mail: \href{mailto:iske0001@e.ntu.edu.sg}{iske0001@e.ntu.edu.sg}\par}
\vskip 0.5em
{\normalsize\textsuperscript{2}\,\textit{Cranfield University, United Kingdom.} E-mail: \href{mailto:leonard.felicetti@cranfield.ac.uk}{leonard.felicetti@cranfield.ac.uk}\par}
\vskip 0.5em
{\normalsize\textsuperscript{3}\,\textit{Istanbul Technical University, T\"{u}rkiye.} E-mail: \href{mailto:gokturkk19@itu.edu.tr}{gokturkk19@itu.edu.tr}\par}
\vskip 0.5em
{\normalsize\textsuperscript{4}\,\textit{Magdrive Ltd, United Kingdom.} E-mail: \href{mailto:dtellett@magdrivespace.com}{dtellett@magdrivespace.com}, \href{mailto:adam@magdrivespace.com}{adam@magdrivespace.com}\par}
\vskip 0.5em
{\normalsize\textsuperscript{*}\,Corresponding author\par}
\vskip 1em
{\small\itshape Preprint of a paper accepted for presentation at the 77th International Astronautical Congress (IAC 2026), Antalya, T\"{u}rkiye, 5--9 October 2026.\par}
\end{center}
\vskip 1em
\begin{center}
\textbf{Abstract}
\end{center}
\noindent
Multi-rendezvous missions for active debris removal and in-orbit servicing incur prohibitive propellant costs when their targets differ in the orientation of their orbital planes, and classical mission design treats the nodal precession induced by Earth's oblateness as a perturbation to be cancelled. This paper presents a drift-aware trajectory optimisation framework that instead exploits that precession as a mission-design resource, benchmarked against the published winning solution to the European Space Agency's Kessler Run trajectory-optimisation competition as a trusted reference. Three methodological contributions are made. First, an enhanced drift orbit shapes the size, shape, and inclination of an intermediate orbit to tune the differential precession rate, substantially cheaper than the altitude-only designs of prior work in the Sun-synchronous regime, where inclination is the dominant lever. Second, a global per-leg time-budget optimiser makes explicit and resolves the coupling between drift-time allocation and downstream nodal geometry that defeats greedy allocation. Third, a sequence-recovery test independently re-derives a removal order consistent with the published reference from the debris catalogue alone. Applied to the full debris campaign on the published mission partition under the competition's transfer-time rules, the framework closely matches the published benchmark cost using only analytical transfer models and minutes of single-core computation. A cross-validation anchored to a widely used flight-dynamics tool quantifies the fidelity limits of the analytic model and bounds the cost of literal execution, and the constructive nature of the algorithm supports rapid in-orbit replanning after a missed manoeuvre.
\par
\vskip 1.5ex
\noindent\textbf{Keywords:} multi-rendezvous trajectory, debris removal, RAAN drift, $J_2$ precession, GTOC9, set-cover backbone, space tug, drift orbit
\vskip 2.5ex
}]

\section*{Nomenclature}
\noindent\small
\begin{tabular}{@{}l@{\;}l@{}}
$\mathcal{D}$              & Target catalog $\{d_1,\ldots,d_M\}$ \\
$n,\,N_i$                  & Mission count; targets visited per mission \\
$J,\; C_i$                  & Total / per-mission cost (MEUR) \\
$c_i,\;\alpha$             & Launch base cost; quadratic mass penalty \\
$m_{dry},m_p,m_{de}$       & Dry, propellant, deorbit-kit mass (kg) \\
$I_{sp},\; g_0$             & Specific impulse; standard gravity \\
$\Delta V$                 & Impulsive velocity change (m/s) \\
$\bm{\kappa}$              & Keplerian element set $\{a,e,i,\Omega,\omega,M\}$ \\
$\mu,\,J_2,\,r_{eq}$       & Earth grav.\ parameter, zonal harmonic, radius \\
$n_o,\,p$                  & Mean motion $\sqrt{\mu/a^3}$; semi-latus rectum \\
$\dot{\Omega},\dot{\omega}$ & RAAN and AoP precession rates \\
$\bm{T},\;\bm{\sigma}$     & Per-leg time-budget vector; target sequence \\
$\Delta t_{wait},\,t_w$    & Drift waiting time; dwell at each target \\
$a_d,e_d,i_d$              & Drift-orbit semi-major axis, eccentricity, inclination \\
$\Delta t_R,\,\Delta t_M$  & Per-leg time cap; per-mission gap (GTOC9) \\
\end{tabular}
\normalsize

\section*{Abbreviations}
\noindent\small
\begin{tabular}{@{}l@{\;\;}l@{}}
\textbf{ACO}    & Ant Colony Optimization \\
\textbf{ADR}    & Active Debris Removal \\
\textbf{DE}     & Differential Evolution \\
\textbf{ESA}    & European Space Agency \\
\textbf{GMAT}   & General Mission Analysis Tool \\
\textbf{GTOC}   & Global Trajectory Optimisation Competition \\
\textbf{IOS}    & In-Orbit Servicing \\
\textbf{JPL}    & Jet Propulsion Laboratory \\
\textbf{LEO}    & Low Earth Orbit \\
\textbf{MEUR}   & Million Euro \\
\textbf{MJD2000} & Modified Julian Date (epoch 2000-01-01) \\
\textbf{NN}     & Nearest Neighbour \\
\textbf{NUDT}   & National Univ.\ of Defense Technology \\
\textbf{P-ACO}  & Population-based ACO \\
\textbf{RAAN}   & Right Ascension of the Ascending Node \\
\textbf{SCP}    & Sequential Convex Programming \\
\textbf{SNOPT}  & Sparse Nonlinear Optimizer \\
\textbf{SSO}    & Sun-Synchronous Orbit \\
\textbf{TSP}    & Travelling Salesman Problem \\
\end{tabular}
\normalsize

\section{Introduction}
\label{sec:introduction}

The accelerating growth of orbital debris threatens long-term space
sustainability. With over 36\,000 tracked objects~\cite{b_esa_discosweb}
and the cascading-collision risk identified by
Kessler~\cite{b_kessler1978}, scalable active debris removal (ADR) is
required, and single-target missions
are economically prohibitive at the necessary cadence. Multi-rendezvous
architectures (a single spacecraft visiting many targets sequentially) offer
a path to cost effectiveness, but couple $\mathcal{NP}$-hard combinatorial
sequencing~\cite{b_izzo2015} with continuous trajectory optimization under
$J_2$-perturbed orbital mechanics.

The 9th Global Trajectory Optimisation Competition (GTOC9)~\cite{b_gtoc9,b_izzo_gtoc9}
has become the canonical benchmark dataset for this class of problem and
provides a controlled, fully-documented environment in which alternative
methods can be evaluated against published reference solutions. The published
JPL solution~\cite{b_jpl} removed all 123 Sun-synchronous debris in 10
missions at a submitted competition score of 731.28~MEUR using a
$\sim$290-million-row database of semi-analytic six-impulse
$J_2$-drift-exploiting transfers,
branch-and-bound sequencing, ant colony
optimization~\cite{b_dorigo1997}, and SNOPT-based multiple-shooting
refinement with up to ten impulses per revolution~\cite{b_snopt}.
Additional published reference solutions from
NUDT~\cite{b_nudt} (786.21~MEUR, 12 missions),
Tsinghua~\cite{b_tsinghua} (829.58~MEUR), DLR~\cite{b_dlr}
(949.85~MEUR), and Strathclyde~\cite{b_strathclyde} (918.98~MEUR, 14
launches, 6th of 69 entries, via beam search and a continuous ``Time
Shuffler'' encoding of the campaign timeline) share a common
structural observation: optimal debris sequences exhibit monotonically
shifting RAAN, a ``RAAN walk''~\cite{b_izzo2015}, because $J_2$-induced
precession yields differential drift rates of 0.2--1.2~deg/day across
the catalogue. We treat these published solutions, and JPL's in
particular, as trusted reference baselines against which the analytical
drift-aware framework developed below is benchmarked for methodological
validation.

Rather than treating $J_2$ as a disturbance to be cancelled, the drift-aware
paradigm exploits nodal precession as a propulsive substitute: small
shape-changing maneuvers ($\sim$100~m/s) modify the local RAAN drift rate by
$\sim$0.1~deg/day, and the required RAAN change accrues passively during
coasting. Three gaps remain. \emph{First}, existing drift orbit models vary
only the semi-major axis $a_d$, neglecting eccentricity and inclination,
even though all three orbital elements appear in the $J_2$ rate
equation~\eqref{eq:raan_rate} and in the SSO regime inclination dominates
because $|\cos i|\!\approx\!0.14$ at $i\!\approx\!98^\circ$. \emph{Second},
the \emph{sequence--timeline coupling}, whereby each leg's drift time shifts
downstream RAAN geometry through accumulated elapsed time, has so far been
treated only stochastically, discretely, locally, or in a decoupled second
stage: JPL allocated time over a fixed sequence via a nearly-convex QP and
evolved discrete per-edge time codes as genes of a genetic
algorithm~\cite{b_jpl}, NUDT's hybrid-encoding GA jointly evolved integer
sequences and real per-leg durations~\cite{b_nudt}, Tsinghua refined arrival
epochs by PSO only after freezing the sequence~\cite{b_tsinghua}, DLR
redistributed leg times locally under a frozen per-mission
total~\cite{b_dlr}, and Strathclyde's Time Shuffler re-encoded the campaign
as a continuous vector of departure times searched by a memetic
algorithm~\cite{b_strathclyde}; none provides a deterministic, global,
continuous per-leg time allocation coupled to a closed-form analytic drift
model. \emph{Third},
the relationship between the GTOC9 quadratic cost function and per-mission
$\Delta V$ has not been systematically exploited.

This paper makes three methodological contributions. First, an enhanced three-parameter drift orbit (C$^{+}$) that simultaneously optimises $(a_d,e_d,i_d)$ and reduces per-leg transfer $\Delta V$ by up to ${\sim}40\%$ on favorable moderate-gap legs relative to altitude-only designs (leg-dependent, and modest at the mission-sequence total; Sec.~\ref{sec:methods}). Second, explicit treatment and resolution of the sequence--timeline coupling whereby each leg's drift time shifts downstream RAAN geometry through accumulated elapsed time, which the two-phase global time-budget optimiser closes by up to an order of magnitude (Sec.~\ref{sec:methods}). Third, a full-campaign GTOC9 validation in which JPL's published mission partition is adopted as a controlled benchmark: applied to all 123 debris on this partition under the framework's GTOC9-referenced per-leg cap (30~days on the \emph{transfer} segment, with the 5-day dwell added separately, a reading up to five days more permissive than the strict arrival-to-arrival $\Delta t_R$; Sec.~\ref{sec:problem}), the framework achieves 742.65~MEUR, remaining within about one and a half percent ($+1.55\%$) of the published JPL reference, and matches or beats the published reference on three of the ten missions (Sec.~\ref{sec:results}). The set-cover problem of \emph{jointly} discovering a mission partition and the per-mission sequences from scratch via a Beam P-ACO backbone~\cite{b_simoes2017} is the natural follow-on and is left to future work; this work intentionally isolates the trajectory-optimization layer from the combinatorial mission-partitioning problem by adopting JPL's published partition purely as a controlled benchmark reference. The objective is methodological validation rather than competitive replication.

This paper is the technical/astrodynamics conference summary of a larger
body of work. Two companion papers extend the present content. The full
methodological development, the GMAT-anchored propagator benchmark, and a 6-scenario
investigation of the per-leg time-budget coupling appear in the journal
manuscript~\cite{iskender_journal2026}. The commercial / operator-facing
framing of the same methodology, packaged as the \emph{Space Postman}
mission-design platform and the \emph{ATLAS} multi-purpose tug, appears
in the companion IAC paper~\cite{iskender_iac_e6} in the IAF Businesses
and Innovation Symposium. The complementary close-proximity
rendezvous-and-docking problem has been treated thoroughly, both
analytically and experimentally, in the authors' prior
work~\cite{iskender2019dual,burak2020phd}; the present paper focuses on
the upstream multi-rendezvous sequencing and per-leg transfer-cost
optimisation rather than the close-proximity terminal phase.

\section{Problem Formulation}
\label{sec:problem}

Let $\mathcal{D}=\{d_1,\ldots,d_M\}$ denote the target catalog of $M$
debris objects, each defined by an osculating Keplerian state
$\bm{\kappa}_k = \{t_{0,k}, a_k, e_k, i_k, \Omega_{0,k},
\omega_{0,k}, M_{0,k}\}$ at reference epoch $t_{0,k}$. A
multi-rendezvous mission visits an ordered subset
$\bm{\sigma} = (d_{\sigma_1}, d_{\sigma_2}, \ldots, d_{\sigma_N})
\subseteq \mathcal{D}$ of $N$ targets within a per-leg transfer-time
budget vector $\bm{T} = (T_1, \ldots, T_{N-1})$. A full GTOC9
\emph{campaign} partitions $\mathcal{D}$ into $n$ such missions
$\{\bm{\sigma}^{(1)},\ldots,\bm{\sigma}^{(n)}\}$ covering all 123 debris
exactly once.

\subsection{Decision variables}

For each mission $i \in \{1,\ldots,n\}$ the planner chooses:
(i) the visitation sequence $\bm{\sigma}^{(i)}$, an $N_i$-permutation of
its target subset;
(ii) the per-leg time budgets $\bm{T}^{(i)} \in [T_{\min},T_{\max}]^{N_i-1}$,
where $T_{\min} \geq t_w + 1$~d (dwell-time floor) and $T_{\max} = \Delta t_R$
(GTOC9 per-leg cap);
(iii) the per-leg transfer method $m_k \in \{A,B,C,C^{+}\}$
(Sec.~\ref{sec:methods}) selecting between direct combined plane change,
natural-wait, single-parameter drift orbit, and three-parameter drift orbit;
and (iv) the global mission start epoch $t_{0}^{(i)}$, taken from JPL's
published schedule in this work.

\subsection{Cost function (PF1)}

Following the GTOC9 specification~\cite{b_izzo_gtoc9} the total campaign
cost is
\begin{equation}
J = \sum_{i=1}^{n} C_i
  = \sum_{i=1}^{n}\!\left[\,c_i + \alpha\,(m_{0,i}-m_{dry})^2\,\right] ,
\label{eq:cost}
\end{equation}
where $C_i$ is the per-mission cost in MEUR, $c_i = 55$~MEUR is the launch
base cost, $\alpha = 2\times 10^{-6}$~MEUR/kg$^2$ penalises heavier
spacecraft, $m_{dry} = 2000$~kg is the dry mass, and the initial mass is
\begin{equation}
m_{0,i} = m_{dry} + N_i\,m_{de} + m_{p,i} ,
\label{eq:mass}
\end{equation}
with $m_{de} = 30$~kg the deorbit-kit mass and $m_{p,i}$ the propellant
mass obtained from the inverse Tsiolkovsky relation
\begin{equation}
m_{p,i} = m_{f,i}\!\left[\exp\!\left(\tfrac{\Delta V_i}{I_{sp}\,g_0}\right)
        -1\right] ,
\quad I_{sp} = 340~\text{s}.
\label{eq:tsiolkovsky}
\end{equation}

The quadratic mass penalty in~\eqref{eq:cost} couples $\Delta V$ to cost
non-linearly, so a sequence with lower total $\Delta V$ does not
necessarily minimise $J$.

\subsection{Constraints}

\begin{align}
T_k & \leq T_{\max}, \quad k=1,\ldots,N_i-1
   \label{eq:c_dtR}\\
\tau_{i+1}-\tau_i & \geq \Delta t_M
   \label{eq:c_dtM}\\
\bm{\sigma}^{(i)} \cap \bm{\sigma}^{(j)} & = \emptyset, \quad \forall\, i \neq j
   \label{eq:c_cover}\\
\bigcup_{i=1}^{n} \bm{\sigma}^{(i)} & = \mathcal{D}
   \label{eq:c_complete}\\
m_{p,i} & \leq m_{p,\max}
   \label{eq:c_mp}\\
\tau_i & \in [t_{\min},\,t_{\max}]
   \label{eq:c_epoch}
\end{align}

Equation~(\ref{eq:c_dtR}) is the per-leg transfer-time cap,
$\tau_i$ in~(\ref{eq:c_dtM}) is the start epoch of mission $i$ and the
inequality enforces the mandatory inter-mission gap, $t_w$ is the
dwell at each target (subsumed into the per-leg accounting),
(\ref{eq:c_cover})--(\ref{eq:c_complete}) make the mission partition a
disjoint and complete cover of $\mathcal{D}$, (\ref{eq:c_mp}) is the
single-launch propellant-tank capacity from the GTOC9 spacecraft
model~\cite{b_izzo_gtoc9}, and (\ref{eq:c_epoch}) is the benchmark mission
timeline. A convention note on~(\ref{eq:c_dtR}) is required for
reproducibility. GTOC9 measures the per-leg cap $\Delta t_R$
\emph{arrival-to-arrival}: the mandatory $\geq$5-day dwell at the
departed target sits inside the 30-day budget, so the effective
transfer window under the strict competition reading is at most
25~days per leg~\cite{b_izzo_gtoc9}. The runs labelled GTOC9-compliant
in this paper apply the 30-day cap to the transfer segment itself, with
the dwell added separately in the elapsed-time propagation; this
reading admits up to five more days per leg (up to 35~days
arrival-to-arrival) than the strict arrival-to-arrival convention, and
is therefore \emph{more permissive} than the competition rule.
Conversely, JPL's transfer database was
restricted to transfers of at most 25~days by
construction~\cite{b_jpl}, so the $\leq$30-day optimiser used here
searches a slightly larger per-leg feasible set than the reference
solution it is compared against. As the conservative counterpart, a
strictly arrival-to-arrival compliant re-run that caps the transfer
segment at 25~days raises the Mission-10 globally-optimised cost to
1\,525~m/s ($+7.2\%$ against the JPL reference), which we quote as the
conservative figure alongside the more permissive 30-day-transfer
values reported below. Numerical values for $T_{\max},\Delta t_M,t_w,m_{p,\max},
m_{dry},m_{de},I_{sp},c_i,\alpha$, and the epoch window
$[t_{\min},t_{\max}]$ are summarised in
Table~\ref{tab:pf_params}.

\begin{table}[h!]
\caption{Problem-formulation parameters (GTOC9 specification~\cite{b_izzo_gtoc9}).}
\label{tab:pf_params}
\centering
\footnotesize
\setlength{\tabcolsep}{4pt}
\begin{tabular}{@{}l@{\hspace{0.4em}}l@{\hspace{0.4em}}c@{}}
\toprule
\textbf{Symbol} & \textbf{Quantity} & \textbf{Value} \\
\midrule
\multicolumn{3}{l}{\emph{Time constraints}} \\
$T_{\max}$       & Per-leg transfer-time cap        & 30~d \\
$\Delta t_M$     & Minimum inter-mission gap        & 30~d \\
$t_w$            & Dwell at each target             & $\geq$5~d \\
$[t_{\min},t_{\max}]$ & Epoch window (MJD2000)       & $[23467,\,26419]$ \\
\midrule
\multicolumn{3}{l}{\emph{Spacecraft model}} \\
$m_{dry}$        & Dry mass                          & 2000~kg \\
$m_{p,\max}$     & Propellant-tank capacity          & 5000~kg \\
$m_{de}$         & Per-debris de-orbit-kit mass      & 30~kg \\
$I_{sp}$         & Specific impulse                  & 340~s \\
$g_0$            & Standard gravity                  & 9.80665~m/s$^2$ \\
\midrule
\multicolumn{3}{l}{\emph{Cost model}} \\
$c_i$            & Base launch cost per mission      & 45--55~MEUR (55 used) \\
$\alpha$         & Quadratic mass penalty            & $2\!\times\!10^{-6}$~MEUR/kg$^2$ \\
\midrule
\multicolumn{3}{l}{\emph{Earth / dynamics}} \\
$\mu$            & Gravitational parameter           & $398{,}600.4418\times 10^9$ m$^3$/s$^2$ \\
$J_2$            & Second zonal harmonic             & $1.08262668\!\times\!10^{-3}$ \\
$r_{eq}$         & Equatorial radius                 & $6{,}378{,}137$~m \\
\bottomrule
\end{tabular}
\end{table}

\subsection{Reduced problem and scope}

Combining \eqref{eq:cost}--\eqref{eq:c_epoch} the joint set-cover and
trajectory-optimisation problem is
\begin{align}
\min_{\{\bm{\sigma}^{(i)},\bm{T}^{(i)},m_k^{(i)},\tau_i\}}
  \quad & J \quad \text{[\,Eq.\,(\ref{eq:cost})\,]} \nonumber\\
\text{s.t.}
  \quad & \text{(\ref{eq:c_dtR})--(\ref{eq:c_epoch})}, \nonumber\\
        & \text{transfer dynamics of Sec.~\ref{sec:dynamics}}.
\label{eq:full_problem}
\end{align}
Problem~\eqref{eq:full_problem} is $\mathcal{NP}$-hard~\cite{b_izzo_gtoc9}:
the disjoint set-cover sub-problem (\ref{eq:c_cover}),(\ref{eq:c_complete})
contains as a special case one of Karp's 21 $\mathcal{NP}$-complete
problems~\cite{b_karp1972}, and the per-leg trajectory sub-problem under
(\ref{eq:c_dtR}) is itself a non-convex continuous program. The present
paper adopts JPL's published partition $\{\bm{\sigma}^{(i)}\}$ as a
controlled benchmark reference to isolate the trajectory-optimization
contribution (decisions (ii),(iii) above) from the orthogonal
combinatorial set-cover decision, leaving joint set-cover discovery to
the journal companion~\cite{iskender_journal2026} and a Beam-P-ACO
follow-on~\cite{b_simoes2017}.

\subsection{Auxiliary objective (PF2)}

An auxiliary $\Delta V$-only objective decouples the trajectory
optimisation from the mass model:
\begin{equation}
\Delta V_{\text{total}}
  = \sum_{i=1}^{n}\sum_{k=1}^{N_i-1}
    \Delta V_k^{*}\!\left(d_{\sigma_k},d_{\sigma_{k+1}},
    t_{\text{elapsed}}^{(k)}(\bm{T})\right) ,
\label{eq:pf2}
\end{equation}
where $\Delta V_k^{*}$ is the cheapest of methods $\{A,B,C,C^{+}\}$ for
leg $k$ at elapsed time $t_{\text{elapsed}}^{(k)}$. PF2 enables direct
comparison with $\Delta V$-only literature results that do not report
mission cost. The framework reports results under both PF1 and PF2 in
Sec.~\ref{sec:results}.

\subsection{Role of secular precession in the cost}
\label{subsec:precession_role}

Both PF1 and PF2 depend on $\Delta V_i(\bm{T},\bm{m})$, and through it on
the inter-target RAAN gap at elapsed time. The $J_2$-induced secular
precession rate $\dot{\Omega}(a,e,i) = -\tfrac{3}{2}J_2(r_{eq}/p)^2 n_o \cos i$
(derived in Sec.~\ref{sec:dynamics}) enters the formulation in two
places. First, the differential rate $\dot{\Omega}_{tgt}-\dot{\Omega}_{src}$
of the source/target pair closes the inter-target RAAN gap passively as
the elapsed time accumulates; this is what makes
the wait-for-drift method~B cheaper than the direct plane change~A for
any non-zero allowed transfer time. Second, the three-parameter drift
orbit~$C^{+}$ chooses an intermediate orbit $(a_d,e_d,i_d) \neq
(a_{src},e_{src},i_{src})$ with a deliberately different
$\dot{\Omega}$, so the gap-closing rate is \emph{amplified}, typically by
a factor of three to five for a few tens of m/s of $(a_d,e_d,i_d)$
insertion overhead (Sec.~\ref{sec:methods}). In the SSO regime $|\cos i| \!\approx\! 0.14$ at
$i \!\approx\! 98^{\circ}$, so a 2$^{\circ}$ change in $i_d$ alters
$|\cos i|$ by $\sim$25\%, against $\sim$4\% per 200~km altitude
change~\cite{iskender_journal2026}; the inclination knob is therefore
the dominant lever, and $\dot{\Omega}$ is exploited as a resource rather
than cancelled as a disturbance.

The argument-of-perigee rate
$\dot{\omega} = \tfrac{3}{4}J_2(r_{eq}/p)^2 n_o(5\cos^2 i - 1)$ is reported
alongside $\dot{\Omega}$ in the dynamics but is \emph{not} propagated
through the transfer-cost model, consistent with the GTOC9
specification's explicit waiver of the velocity contribution of
$\dot{\Omega}$ and $\dot{\omega}$~\cite{b_izzo_gtoc9}; the consequences of
that simplification are characterised against numerical $J_2$ truth in
Sec.~\ref{sec:gmat} and the journal companion~\cite{iskender_journal2026}.

\subsection{Problem class and solver}
\label{subsec:solver_class}

The full design problem~\eqref{eq:full_problem} is a non-convex,
non-separable mixed-integer programme: continuous time-budget vectors
$\bm{T}$, discrete method assignments $\bm{m}$, discrete sequences
$\bm{\sigma}$, and integer mission counts $n$ enter a quadratic-mass cost
through an exponential propellant relation~\eqref{eq:tsiolkovsky}. The
$\mathcal{NP}$-hard set-cover sub-problem
(\ref{eq:c_cover})--(\ref{eq:c_complete}) is fixed to JPL's published
partition in this paper, reducing the residual problem to the
per-mission mixed-integer programme (decision vectors $\bm{T},\bm{m}$;
the per-leg time-budget vector and the discrete method labels) of
Sec.~\ref{subsec:nn2opt}; the formal statement is in the journal
companion~\cite{iskender_journal2026}. The per-mission MIP is non-convex (in $\bm{T}$
because of the gap-modulated transfer cost) and non-separable (each
$T_j$ shifts the elapsed time on every downstream leg). It is not
amenable to gradient-based or convex-relaxation solvers, so a
two-phase metaheuristic is used: Differential
Evolution~\cite{b_storn1997} (SciPy \texttt{scipy.optimize.}\linebreak[1]\texttt{differential\_evolution},
population 20, $F=0.5$, $CR=0.7$, $\leq$100 generations, fixed seed~42)
on the $\{A,B\}$ relaxation for speed, followed by a deterministic
coordinate-descent refinement over a discrete $T_k$ grid with the full
method catalog $\{A,B,C,C^{+}\}$ available
(Sec.~\ref{subsec:nn2opt}). For target sequencing within a mission, a
greedy nearest-neighbour construction followed by 2-opt local search is
used; the Beam P-ACO sequencer~\cite{b_simoes2017} is invoked only when
the partition is to be discovered jointly. All optimisation runs single-core
on a laptop-class CPU: $\sim$27~minutes for the full 123-debris campaign,
underscoring the computational accessibility of the analytical formulation.

\section{Orbital Dynamics and Drift Exploitation}
\label{sec:dynamics}

\subsection{$J_2$-Perturbed Secular Precession}

Between maneuvers, the spacecraft and debris evolve under Keplerian
dynamics perturbed by Earth's $J_2$ harmonic. The secular precession
rates of the ascending node and argument of perigee are~\cite{b_izzo_gtoc9}:
\begin{align}
\dot{\Omega} &= -\frac{3}{2}J_2\!\left(\frac{r_{eq}}{p}\right)^{\!2}\! n_o\cos i ,
\label{eq:raan_rate} \\
\dot{\omega} &= \frac{3}{4}J_2\!\left(\frac{r_{eq}}{p}\right)^{\!2}\!
n_o(5\cos^2\!i - 1) ,
\label{eq:aop_rate}
\end{align}
where $n_o=\sqrt{\mu/a^3}$ and $p=a(1-e^2)$. The osculating elements at
epoch~$t$ are propagated linearly: $\Omega(t)=\Omega_0+\dot{\Omega}\,\Delta t$,
$\omega(t)=\omega_0+\dot{\omega}\,\Delta t$, $M(t)=M_0+n_o\,\Delta t$.

\begin{remark}[Cost-model consumption of secular rates]
\label{rem:secular}
The trajectory cost model in this paper consumes only the secular RAAN rate
$\dot{\Omega}$ from~\eqref{eq:raan_rate}. The argument-of-perigee rate
$\dot{\omega}$ is reported alongside $\dot{\Omega}$ in the dynamics but is
not propagated through the transfer-cost evaluation; the secular evolution
of $M$ is treated implicitly, with true-anomaly phasing at rendezvous
absorbed into the transfer-time budget rather than enforced as an explicit
equality. These choices are consistent with the GTOC9 specification, which
explicitly waives the velocity contribution of $\dot{\Omega}$ and $\dot{\omega}$
on the grounds that it ``removes complexity from the equations without
introducing any significant change on the search space
landscape''~\cite{b_izzo_gtoc9}, and they are appropriate for the tangential
Hohmann-class transfers used here, since burn placement at perigee/apogee
absorbs AoP rotation geometrically. The systematic error introduced is
quantified empirically in Sec.~\ref{sec:gmat} (GMAT-anchored FPROX
propagator benchmark + per-leg plane-acquisition verification of the
analytic plan).
\end{remark}

The RAAN difference between two orbits evolves linearly with elapsed time.
A spacecraft reaches a target requiring RAAN change $\Delta\Omega$ by
coasting on a drift orbit at $(a_d,e_d,i_d)$ for
\begin{equation}
t_{drift} = \frac{\Delta\Omega}{\dot{\Omega}(a_d,e_d,i_d) - \dot{\Omega}_{tgt}} .
\label{eq:tdrift}
\end{equation}

\subsection{Three-Parameter Drift Orbit (C+)}

Standard GTOC9 approaches vary only~$a_d$ in~\eqref{eq:tdrift}. From
\eqref{eq:raan_rate}, $\dot{\Omega}$ depends on three independent elements
through $p=a(1-e^2)$ and $\cos i$. For SSO ($i\!\approx\!96\text{--}102^\circ$),
inclination is the dominant lever: at $i=98^\circ$,
$|\cos i|\!\approx\!0.14$, so a $2^\circ$ change in~$i$ alters
$|\cos i|$, and hence the nodal rate, by~$\sim$25\%, while a 200-km
altitude change at $\sim$750~km alters $\dot{\Omega}$ by only~$\sim$4\%. Table~\ref{tab:three_knobs} compares the
three knobs side-by-side.

\begin{table*}[t]
\caption{The three knobs of the enhanced drift orbit transfer in the GTOC9
SSO regime. Inclination dominates because $|\cos i|\!\approx\!0.14$ at
$i\!\approx\!98^\circ$, so a small change in~$i$ produces a disproportionately
large fractional change in $\dot{\Omega}$.}
\label{tab:three_knobs}
\centering
\footnotesize
\begin{tabular}{@{}llcc@{}}
\toprule
\textbf{Knob} & \textbf{Mechanism in $\dot{\Omega}$} & \textbf{Typical leverage} & \textbf{$\Delta V$ cost} \\
\midrule
$a_d$ -- altitude     & changes $p$ and $n_o=\sqrt{\mu/a^3}$       & $-$200~km $\to +0.05^\circ$/day & $\sim$200--300~m/s \\
$e_d$ -- eccentricity & reduces $p$ at fixed $a$                 & $e\!=\!0\!\to\!0.05$: $-$0.25\% in $p$ & $\sim$10--30~m/s \\
$i_d$ -- inclination  & modifies $|\cos i|$ (dominant)           & $\Delta i\!=\!2^\circ\to$ 25\% nodal-rate change & $\sim$100~m/s \\
\midrule
$(a_d,e_d,i_d)$ -- C+ & coupled optimisation of all three        & up to ${\sim}40\%$ per leg vs.\ $a_d$ only & weighted optimal \\
\bottomrule
\end{tabular}
\end{table*}

\begin{figure*}[t]
\centering
\includegraphics[width=\textwidth]{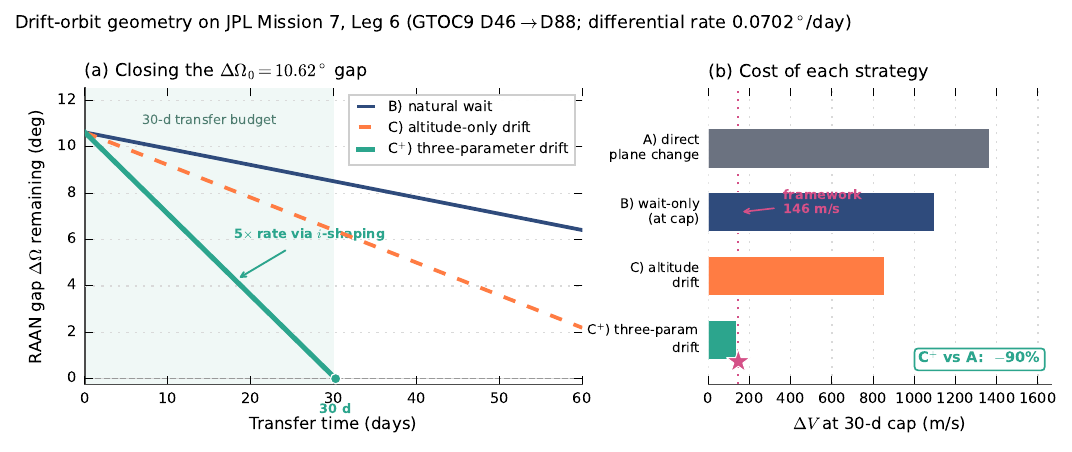}
\caption{Three-parameter drift-orbit concept on JPL Mission~7 Leg~6
(GTOC9 D46$\to$D88): (a)~closure of the RAAN gap and (b)~the resulting
$\Delta V$ at the 30-d cap for the direct (A), wait-only (B), altitude-only
(C) and three-parameter drift (C$^+$) strategies; the magenta star marks the
framework's full numerical result. The A/B/C/C$^+$ curves are drawn from an
\emph{assumed} effective-rate model ($2\times$/$5\times$ rate amplification
with a fixed $+30$/$+40$~m/s insertion overhead) \emph{calibrated against}
the optimiser rather than produced by it; the ``$\sim$5$\times$
amplification'' and ``$\sim$40~m/s insertion'' are therefore model inputs,
not findings. Only the magenta star is a direct optimiser output.}
\label{fig:drift_concept}
\end{figure*}

The enhanced method fixes $t_{drift}$ and solves for $(e_d,i_d)$ on a
discrete grid ($e_d\in[0,0.06]$, $i_d\in[i_{src}-3^\circ,i_{src}+3^\circ]$),
with $a_d$ obtained from~\eqref{eq:tdrift} by bisection for each pair. The
entry and exit are each a two-impulse Hohmann-type transfer between a circular
orbit and the elliptical drift orbit, with the combined plane change placed at
whichever of perigee or apogee minimises cost; a drift-orbit leg (C/C$^+$)
therefore uses four impulses in total, versus two for a direct leg~(A). Fig.~\ref{fig:drift_concept} illustrates
the geometry on a representative GTOC9 leg taken from JPL Mission~7 Leg~6
(D46$\to$D88, $\Delta\Omega_0 = 10.62^\circ$, differential rate
$0.0702^\circ$/day, $V_{\text{base}} = 95$~m/s). Panel~(a) shows the RAAN gap
closing under each strategy: the three-parameter drift orbit (C$^+$) amplifies
the natural differential rate $\sim$5$\times$ via inclination shaping and
closes the $10.62^\circ$ gap inside the GTOC9 $\Delta t_R\!\leq\!30$~d/leg
cap, whereas the natural-wait baseline (B) needs $\sim$151~d to converge to
its absolute minimum $V_{\text{base}}\approx 95$~m/s. Panel~(b) converts this
into cost at the binding 30-d cap: the direct combined plane change (A) costs
$\sim$1\,365~m/s, natural wait (B) $\sim$1\,096~m/s, altitude-only drift (C)
$\sim$858~m/s, and C$^+$ only $\sim$136~m/s, a 90\% saving over A. The
framework's full numerical result for this leg ($\sim$146~m/s at $\leq$30~d)
is marked by the magenta star. The campaign-wide generalisation across all 113
legs is reported in Section~\ref{sec:results} (Table~\ref{tab:fig1_methods}
gives the $\Delta V$ summary on this leg, parallel to the journal manuscript's
single-leg analysis on M7~L7).

\begin{figure*}[t]
\centering
\includegraphics[width=\textwidth]{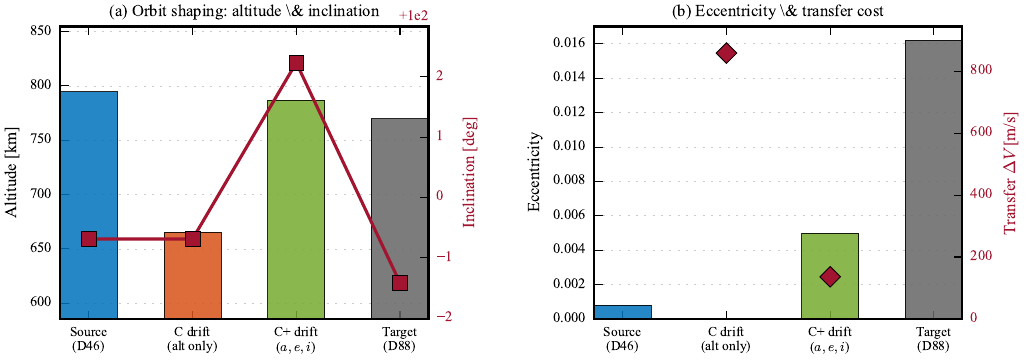}
\caption{Drift-orbit element comparison for M7~L6 (D46$\to$D88, $\leq$30~d
cap): altitude and inclination (a) versus eccentricity and transfer-$\Delta V$
(b) for the single-parameter (C) and three-parameter (C$^+$) drift orbits.}
\label{fig:drift_elements_canonical}
\end{figure*}

Fig.~\ref{fig:drift_elements_canonical} contrasts how the single-parameter
(C) and three-parameter (C$^+$) drift orbits achieve the \emph{same}
$\dot{\Omega}$ on this leg through different elements: C lowers altitude by
129~km, whereas C$^+$ tilts inclination by $+2.92^\circ$ and barely changes
altitude, costing 84\% less $\Delta V$ (136 vs.\ 858~m/s).

\begin{table*}[t]
\caption{Transfer strategies plotted in Fig.~\ref{fig:drift_concept} on
M7~L6 (D46$\to$D88, $\Delta\Omega_0=10.62^\circ$, differential rate
$0.0702^\circ$/day, $V_{\text{base}}=95$~m/s). Analytical $\Delta V$
values are read off an \emph{assumed} effective-rate model (the same
$2\times$/$5\times$ amplification with fixed $+30$/$+40$~m/s insertion as
Fig.~\ref{fig:drift_concept}), calibrated against (not produced by) the
optimiser, at the optimum within the GTOC9 $\leq$30~d/leg cap. The
``$\sim$5$\times$ amplification'' and ``$\sim$40~m/s insertion'' are model
inputs, not measured results; only the $\star$ row (146~m/s) is a direct
optimiser output.}
\label{tab:fig1_methods}
\centering
\footnotesize
\setlength{\tabcolsep}{4pt}
\begin{tabular}{@{}llccc@{}}
\toprule
\textbf{Curve} & \textbf{Strategy} & \textbf{Knobs tuned} & \textbf{$\Delta V$ @ $t_{\text{opt}}\leq 30$\,d} & \textbf{$t_{\text{opt}}$} \\
 & & & [m/s] & [d] \\
\midrule
A) grey dashed line & Direct combined plane change (no drift) & none & 1\,365 & 0 \\
B) blue solid       & Wait at source orbit for natural drift  & wait time $t$ & 1\,096 & 30 \\
C) orange dashed    & Single-parameter drift orbit            & $a_d$ only & 858 & 30 \\
\textbf{C$^+$) green solid} & \textbf{Three-parameter drift orbit}     & $(a_d, e_d, i_d)$ & \textbf{136} & \textbf{30} \\
\midrule
$\star$) magenta star & Present framework numerical ($\leq$30~d cap) & all methods + global $\bm{T}$ & 146 & n/a \\
\bottomrule
\multicolumn{5}{l}{\scriptsize Saving of analytical C$^+$ over B at 30~d: $-$960~m/s ($-$88\%); over C at 30~d: $-$722~m/s ($-$84\%).} \\
\multicolumn{5}{l}{\scriptsize Natural minimum for B (no time cap) is $V_{\text{base}}\approx 95$~m/s at $t\approx 151$~d, outside the GTOC9 30-d cap.} \\
\multicolumn{5}{l}{\scriptsize The framework's numerical $\sim$146~m/s reflects the full method selection and global time-budget optimiser.} \\
\end{tabular}
\end{table*}

\section{Multi-Rendezvous Optimization Pipeline}
\label{sec:methods}

Before the per-leg details, the algorithmic structure deserves a high-level
justification. The multi-rendezvous problem is fundamentally \emph{mixed}: a
discrete, $\mathcal{NP}$-hard target-sequencing decision (a permutation and
set-cover choice that exposes no usable gradient) sits on top of a continuous
per-leg time-budget and transfer-design problem. The resulting objective is
multimodal and non-smooth: the $J_2$ RAAN geometry is periodic and wraps
around, and the sequence--timeline coupling introduced below creates
near-discontinuous cost ``cliffs'' where a downstream leg's plane change blows
up or a leg becomes infeasible. A monolithic gradient-based or
convex-relaxation solver applied to such a landscape either stalls in a local
trap or fails on the non-differentiable boundaries. We therefore adopt a
two-phase scheme that decouples the combinatorial and continuous parts:
Phase~1 fixes the discrete sequence (nearest-neighbour with 2-opt here; Beam
P-ACO at full-catalogue scale), and Phase~2 optimises the continuous per-leg
time budget on that fixed sequence with derivative-free methods (differential
evolution followed by coordinate-descent refinement) that tolerate the
multimodality and non-smoothness. This separation keeps the problem tractable
in minutes of single-core compute while remaining consistent with the
cost levels reported in the published GTOC9 reference solutions.

\subsection{Per-Leg Transfer Methods}

Four transfer strategies are evaluated per leg:

\noindent\textbf{A) Direct.} Combined Hohmann + plane change at the higher
orbit. Zero wait time; highest propellant cost for large RAAN gaps.

\noindent\textbf{B) Wait-for-drift.} Spacecraft coasts in the source orbit
while natural $J_2$ precession closes the RAAN gap (1-day grid search up to
200~days), followed by a direct transfer.

\noindent\textbf{C) Single-parameter drift.} Three-phase maneuver (enter
drift orbit / coast / exit) with altitude-only ($a_d$) optimization.

\noindent\textbf{C+) Three-parameter drift.} Simultaneous $(a_d,e_d,i_d)$
optimization per Sec.~\ref{sec:dynamics}.

\noindent The hybrid method selects $\min(A,B,C^+)$ per leg; since C is the
$e_d\!=\!0$, $i_d\!=\!i_{src}$ special case of C$^{+}$,
$\min(A,B,C^{+})=\min(A,B,C,C^{+})$, and C is evaluated separately only for
the single-knob ablation of Table~\ref{tab:fig1_methods} and
Fig.~\ref{fig:drift_elements_canonical}. Propellant mass
evolves via Tsiolkovsky: $m_f=m_i\exp(-\Delta V/v_e)$ with
$v_e=I_{sp}\,g_0$.

\subsection{Sequence--Timeline Coupling and Global Time Budget}

Each leg's drift or wait time shifts the RAAN of all downstream targets
through accumulated elapsed time. For instance, a 60-day drift on Leg~1 can
triple a downstream RAAN gap from $12^\circ$ to $36^\circ$, turning a
270~m/s transfer into a 4\,600~m/s plane change. This \emph{sequence--timeline
coupling} makes greedy per-leg time allocation catastrophically suboptimal
when drift time is unconstrained.

For brevity, the formal per-mission sub-problem (decision vectors
$\bm{T},\bm{m}$; cost~\eqref{eq:cost}; inverse-Tsiolkovsky propellant mass;
accumulated-elapsed-time recursion as the non-separability source; bounds
$T_k\in[t_w+1,T_{\max}]$ and $m_k\in\{A,B,C,C^{+}\}$; secular RAAN
propagation) is stated as Eqs.~(15)--(22) of the journal
companion~\cite{iskender_journal2026}, which this paper takes as given.

\paragraph{Optimisation algorithm.}
The per-mission sub-problem is a non-convex mixed-integer program
(continuous $\bm{T}$, discrete $\bm{m}$) with a non-separable objective. A two-phase scheme handles it:
(1)~Differential Evolution~\cite{b_storn1997}
(SciPy implementation, population~20, $F=0.5$, $CR=0.7$, $\leq$100
generations, fixed seed~42) over a relaxation that drops $C^{+}$
($\bm{m}\in\{A,B\}^{N_i-1}$ for speed);
(2)~coordinate-descent refinement over
$T_k\in\{0.5,1,2,5,10,16,20,30,45,60\}\cap[t_w+1,\Delta t_R]$
(up to 10 passes, 0.1~m/s threshold) with the full method
catalog $\{A,B,C,C^{+}\}$ available.
Total runtime: 3--10~s per mission on a single core; approximately
27~minutes for the full 123-debris campaign of Sec.~\ref{sec:results}.

\subsection{Sequence Generation: NN + 2-opt}
\label{subsec:nn2opt}

For each mission, a greedy nearest-neighbour construction (starting from
every debris as initial node) followed by 2-opt local search refines up to
$N$ candidate sequences. Crucially, every step uses the accumulated elapsed
time including dwell to propagate RAANs via~\eqref{eq:raan_rate}, so the
construction captures the coupling effect.

\subsection{Set-Cover Backbone (Future Work)}
\label{subsec:setcover}

The GTOC9 problem in its full form is a set-cover variant: the 123 debris
must be \emph{partitioned} into missions, and within each mission a removal
sequence must be chosen. The state-of-the-art combinatorial backbone for
this two-level problem is Beam Population-based Ant Colony Optimization
(Beam P-ACO)~\cite{b_simoes2017}, which combines pheromone learning with
beam-search pruning to generate a pool of $|\mathcal{S}|\!\sim\!31\,000$
candidate missions small enough for the set-cover ILP to remain tractable
for solvers such as SCIP~\cite{b_scip}.
Complementary tree-search formulations using Monte Carlo Tree
Search~\cite{b_hennes_izzo_mcts} and parallel multi-objective optimisation
frameworks such as pagmo~\cite{b_biscani_izzo_pagmo} have been used in
related GTOC-class problems.

\textbf{Scope.} This paper does \emph{not} run Beam P-ACO on the full
catalogue. The full-campaign result of Sec.~\ref{sec:results} adopts the
published JPL mission partition as a controlled benchmark reference and
applies the NN+2-opt pipeline of Sec.~\ref{subsec:nn2opt} together with
the per-mission global time-budget optimiser. Joint partition-and-sequence discovery via Beam P-ACO,
without any prior from the published JPL solution, is the natural
follow-on extension of this framework.

\section{GTOC9 Benchmark and Constants}
\label{sec:gtoc9}

Table~\ref{tab:constants} lists the physical constants and operational
parameters used throughout, taken from the ESA GTOC9 problem
definition~\cite{b_izzo_gtoc9}.

\begin{table}[h!]
\caption{GTOC9 benchmark constants and operational parameters
(ESA official~\cite{b_izzo_gtoc9}).}
\label{tab:constants}
\centering
\footnotesize
\begin{tabular}{@{}lll@{}}
\toprule
\textbf{Parameter} & \textbf{Value} & \textbf{Units} \\
\midrule
\multicolumn{3}{l}{\emph{Cost function}} \\
$\alpha$ & $2.0\!\times\!10^{-6}$ & MEUR/kg$^2$ \\
$c_m/c_M$ & 45 / 55 & MEUR \\
\midrule
\multicolumn{3}{l}{\emph{Spacecraft}} \\
$m_{dry}$ & 2\,000 & kg \\
$m_p$ (max) & 5\,000 & kg \\
$m_{de}$ & 30 & kg \\
$I_{sp}$ & 340 & s \\
$K_{max}$ & 5 & impulses/leg \\
\midrule
\multicolumn{3}{l}{\emph{Operational}} \\
$\Delta t_R$ & 30 & days \\
$\Delta t_M$ & 30 & days \\
$t_w$ & $\geq$5 & days \\
$[t_{min},t_{max}]$ & $[23467, 26419]$ & MJD2000 \\
\midrule
\multicolumn{3}{l}{\emph{Physical constants}} \\
$\mu$ & $3.986004418\!\times\!10^{14}$ & m$^3$/s$^2$ \\
$J_2$ & $1.08262668\!\times\!10^{-3}$ & -- \\
$r_{eq}$ & 6\,378\,137 & m \\
$g_0$ & 9.80665 & m/s$^2$ \\
\bottomrule
\end{tabular}
\end{table}

A critical observation for the validation that follows: debris orbital
elements are catalogued at \emph{different} reference epochs
($t_{ref}\!\approx$ MJD2000 22\,000), while JPL's missions begin at
$t_0\!\approx\!26\,267$~MJD2000, $\sim$4\,200~days later. Each RAAN must
be propagated forward, accumulating~$\sim$4\,200$^\circ$ of precession per
debris. After correct epoch propagation, the 10 objects of JPL Mission~10
cluster within a $\sim$34$^\circ$ RAAN band (257$^\circ$--291$^\circ$). A
na\"ive analysis using reference-epoch RAANs gives per-leg costs of
3\,000--15\,000~m/s, orders of magnitude too high.

\section{Results}
\label{sec:results}

The framework is exercised on three layers of benchmark validation.
Sec.~\ref{subsec:m10} is a single-mission deep-dive on JPL Mission~10
(10 debris), the published reference solution providing a controlled
testbed for per-leg behaviour. Sec.~\ref{subsec:full_campaign} applies the
same pipeline to the full 123-debris GTOC9 campaign on the published JPL
10-mission partition and reports the headline benchmark-validation
result. Sec.~\ref{subsec:other_cases} summarises two side validations on a
non-GTOC9 SSO inspection mission and a commercial space tug, included
to show transferability across mission archetypes.

\subsection{Single-Mission Deep-Dive on Published JPL Mission~10 Sequence}
\label{subsec:m10}

\begin{figure*}[t]
\centering
\includegraphics[width=\textwidth]{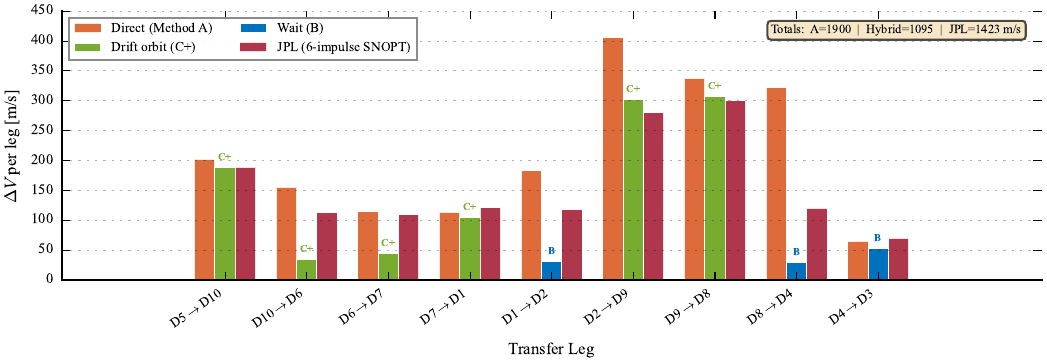}
\caption{Per-leg $\Delta V$ on the published JPL Mission~10 reference
sequence, with labels above each hybrid bar marking the selected
sub-method (A, B, or C$^+$). Bars are shown alongside the published JPL
reference values for direct visual comparison.}
\label{fig:method_comparison}
\end{figure*}

Table~\ref{tab:case2_legs} compares per-leg $\Delta V$ across methods for
the published JPL Mission~10 sequence, used here as a controlled
single-mission benchmark. The direct baseline (A) totals 1\,901~m/s.
The unconstrained hybrid (which violates the GTOC9
$\Delta t_R\!\leq\!30$~d/leg cap and is presented for \emph{sensitivity
analysis only}) achieves 1\,095~m/s but requires a 442-day mission. Under
the $\leq$16~d/leg constraint corresponding to the published JPL
schedule, the per-leg table (evaluated at mission-start RAAN geometry)
totals 1\,652~m/s in 126~days; the sequence-consistent evaluation, in
which each leg's drift shifts the geometry of every later leg, gives the
operative figure of 1\,669~m/s in 143~days ($+17\%$ vs.\ the published
reference; Table~\ref{tab:global_opt}). The per-leg
breakdown in Fig.~\ref{fig:method_comparison} shows that C$^+$ dominates
legs with moderate RAAN gaps ($\sim$1--1.4$^\circ$), while the
natural-wait method~B suffices for the smallest gaps.

\begin{table*}[t]
\caption{Per-leg $\Delta V$ comparison on the published JPL
Mission~10 sequence used as a controlled benchmark.
Hybrid$=\min(A,B,C^+)$. The ``JPL'' column lists the published
reference values. The unconstrained column is a sensitivity result that
violates the GTOC9 $\Delta t_R\!\leq\!30$~d/leg constraint and is
reported only to characterise the algorithm's behaviour absent the
time cap.}
\label{tab:case2_legs}
\centering
\footnotesize
\begin{tabular}{@{}crrrrr@{}}
\toprule
 & & \multicolumn{4}{c}{\textbf{$\Delta V$ [m/s]}} \\
\cmidrule(l){3-6}
\textbf{Leg} & \textbf{$\Delta\Omega$ [$^\circ$]} &
\textbf{A: Direct} & \textbf{Hyb. (unc.)} & \textbf{Hyb. ($\leq$16\,d)} &
\textbf{JPL ref.} \\
\midrule
D5$\to$D10 & $+0.1$ & 203 & 189 & 189 & 189 \\
D10$\to$D6 & $-1.2$ & 156 &  34 & 117 & 113 \\
D6$\to$D7  & $-0.8$ & 115 &  45 & 106 & 110 \\
D7$\to$D1  & $-0.5$ & 114 & 104 & 114 & 121 \\
D1$\to$D2  & $-1.4$ & 183 &  31 & 130 & 118 \\
D2$\to$D9  & $-2.3$ & 406 & 302 & 406 & 280 \\
D9$\to$D8  & $-0.8$ & 338 & 307 & 307 & 300 \\
D8$\to$D4  & $-2.4$ & 322 &  30 & 231 & 121 \\
D4$\to$D3  & $-0.2$ &  66 &  54 &  54 &  70 \\
\midrule
\textbf{Total} & & \textbf{1\,901} & \textbf{1\,095$^\ddagger$} &
\textbf{1\,652} & \textbf{1\,423} \\
\textbf{Mission [d]} & & 50$^\dagger$ & 442 & 126 & 148 \\
\bottomrule
\multicolumn{6}{l}{\scriptsize $^\dagger$Method~A: zero transfer; mission $=$ dwell only.} \\
\multicolumn{6}{l}{\scriptsize $^\ddagger$Sensitivity only: violates GTOC9 $\Delta t_R\!\leq\!30$~d/leg cap.} \\
\multicolumn{6}{l}{\scriptsize Per-leg values at mission-start RAAN geometry (not sequence-consistent);} \\
\multicolumn{6}{l}{\scriptsize \quad totals from unrounded values, columns may not sum exactly (see text).} \\
\end{tabular}
\end{table*}

\paragraph{Three-parameter drift-cost reduction.}
The per-leg saving of C$^+$ over the single-parameter baseline (C,
altitude-only) is strongly convention-dependent and must be quoted with
its measurement recipe. Evaluated under a uniform $\leq$16-d/leg cap with
a 16-d/leg clock (each leg costed independently at that budget), C$^+$
reduces per-leg drift cost by $\sim$35.8\%, $\sim$47.1\%, and $\sim$52.8\%
on the three most favourable Mission-10 legs, up to $\sim$53\%. The
higher single-leg figures reported earlier (43.5\% on D6$\to$D7, 41.8\%
on D7$\to$D1) come from the journal companion's fixed mission-start
geometry ablation under the $\leq$30-d/leg cap and are retained only with
that attribution~\cite{iskender_journal2026}. Either way the gain is
leg-dependent (no benefit on the hardest large-gap legs, and a
mission-sequence-total change within a few percent) and is driven by the
$\cos i$ lever in~\eqref{eq:raan_rate}.

\begin{figure}[t]
\centering
\includegraphics[width=\columnwidth]{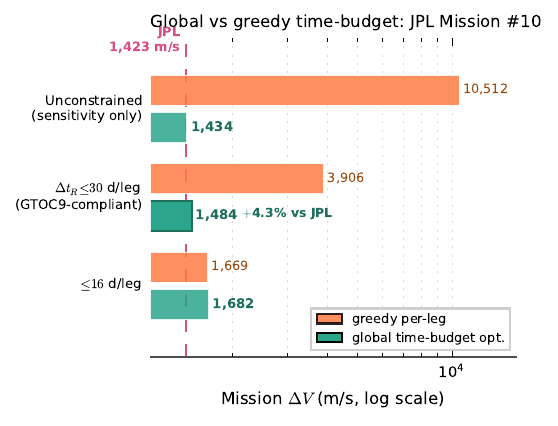}
\caption{Reduction ladder for the global time-budget optimiser on the
published JPL Mission~10 reference sequence: greedy per-leg versus
globally-optimised mission $\Delta V$ across three constraint regimes,
with the published JPL reference value of 1\,423~m/s (dashed) shown for
benchmark context.}
\label{fig:global_opt}
\end{figure}

\paragraph{Global time-budget optimization.}
Table~\ref{tab:global_opt} quantifies the impact of the global optimizer.
Under unconstrained drift (sensitivity only), greedy allocation yields
10\,512~m/s because early long drifts destroy downstream RAAN geometry;
the global optimizer recovers to 1\,434~m/s. Under the GTOC9-compliant
$\leq$30~d/leg constraint, the optimizer drops $\Delta V$ from 3\,906~m/s
(greedy) to 1\,484~m/s, within 4.3\% of the published JPL reference value.
Under the tighter $\leq$16~d/leg cap reported by JPL, the constraint
itself implicitly suppresses coupling and the optimizer offers no further
benefit. Fig.~\ref{fig:global_opt} visualises the reduction across the
three constraint regimes.

\begin{table}[h!]
\caption{Global time-budget vs.\ greedy per-leg allocation on JPL's
Mission~10 sequence (both C+ hybrid; $\Delta V$ in m/s). ``Rel.\ JPL''
is the residual against the published JPL value; the unconstrained row
is sensitivity-only.}
\label{tab:global_opt}
\centering
\footnotesize
\resizebox{\columnwidth}{!}{%
\begin{tabular}{@{}lrrrr@{}}
\toprule
\textbf{Constraint} & \textbf{Greedy} & \textbf{Global} &
\textbf{Rel.\ JPL} & \textbf{Save} \\
\midrule
Unconstrained$^\ddagger$       & 10\,512 & \textbf{1\,434} & $+0.8\%$ & 86\% \\
$\Delta t_R\!\leq\!30$\,d/leg$^\dagger$ &  3\,906 & \textbf{1\,484} & $+4.3\%$ & 62\% \\
$\leq$16\,d/leg                &  1\,669 & 1\,682           & $+18\%$  & $-0.8\%$ \\
\midrule
JPL ref.\ value                & \multicolumn{2}{c}{1\,423} & n/a & n/a \\
\bottomrule
\end{tabular}}
\par\vspace{2pt}\noindent{\scriptsize $^\dagger$ GTOC9-compliant (30-d transfer segment; Sec.~\ref{sec:problem}); $^\ddagger$ violates $\Delta t_R\!\leq\!30$~d/leg, sensitivity only.}
\par\vspace{1pt}\noindent{\scriptsize The unconstrained (1\,434) and 30-d (1\,484) global values are from an archived deep-dive optimiser configuration (maxiter~100, popsize~20, 10 coordinate-descent passes) and will not reproduce bit-for-bit from the current campaign pipeline (maxiter~60, popsize~15, 2 passes), which returns 1\,414~m/s unconstrained and, under the strict 25-d arrival-to-arrival transfer cap, 1\,525~m/s ($+7.2\%$).}
\end{table}

\paragraph{Independent sequence-recovery validation.}
Given only the debris catalogue (orbital elements and reference epochs,
without using the published JPL sequence as a prior), the NN+2-opt
pipeline with Method~B (wait-for-drift) under the $\leq$16~d/leg
constraint independently re-derives a removal sequence
D5$\to$D10$\to$D6$\to$D7$\to$D1$\to$D2$\to$D9$\to$D8$\to$D4$\to$D3 that
is structurally consistent with the published JPL reference sequence,
achieving 1\,831~m/s in 141~days. This recovery serves as a validation
case for the sequencing layer and confirms that the optimal sequence is
driven primarily by $J_2$ precession geometry, not by higher-order
dynamics or fine-grained trajectory phasing.

\subsection{Full 123-Debris Campaign on the Published JPL Mission Partition}
\label{subsec:full_campaign}

The single-mission analysis above establishes per-leg behaviour. We now
apply the same NN+2-opt~+~global-time-budget pipeline to the
\emph{complete} GTOC9 campaign (all 10 missions and all 123 debris) using
the published JPL mission partition as a controlled benchmark reference.
This isolates the trajectory-optimization layer from the orthogonal
combinatorial set-cover decision, which is not addressed in this study and
which we adopt verbatim from the published JPL solution
(Sec.~\ref{subsec:setcover}). The objective is to evaluate the
analytical transfer model and the global time-budget allocator against
a trusted reference solution as a methodological validation exercise.

\begin{figure*}[t]
\centering
\includegraphics[width=\textwidth]{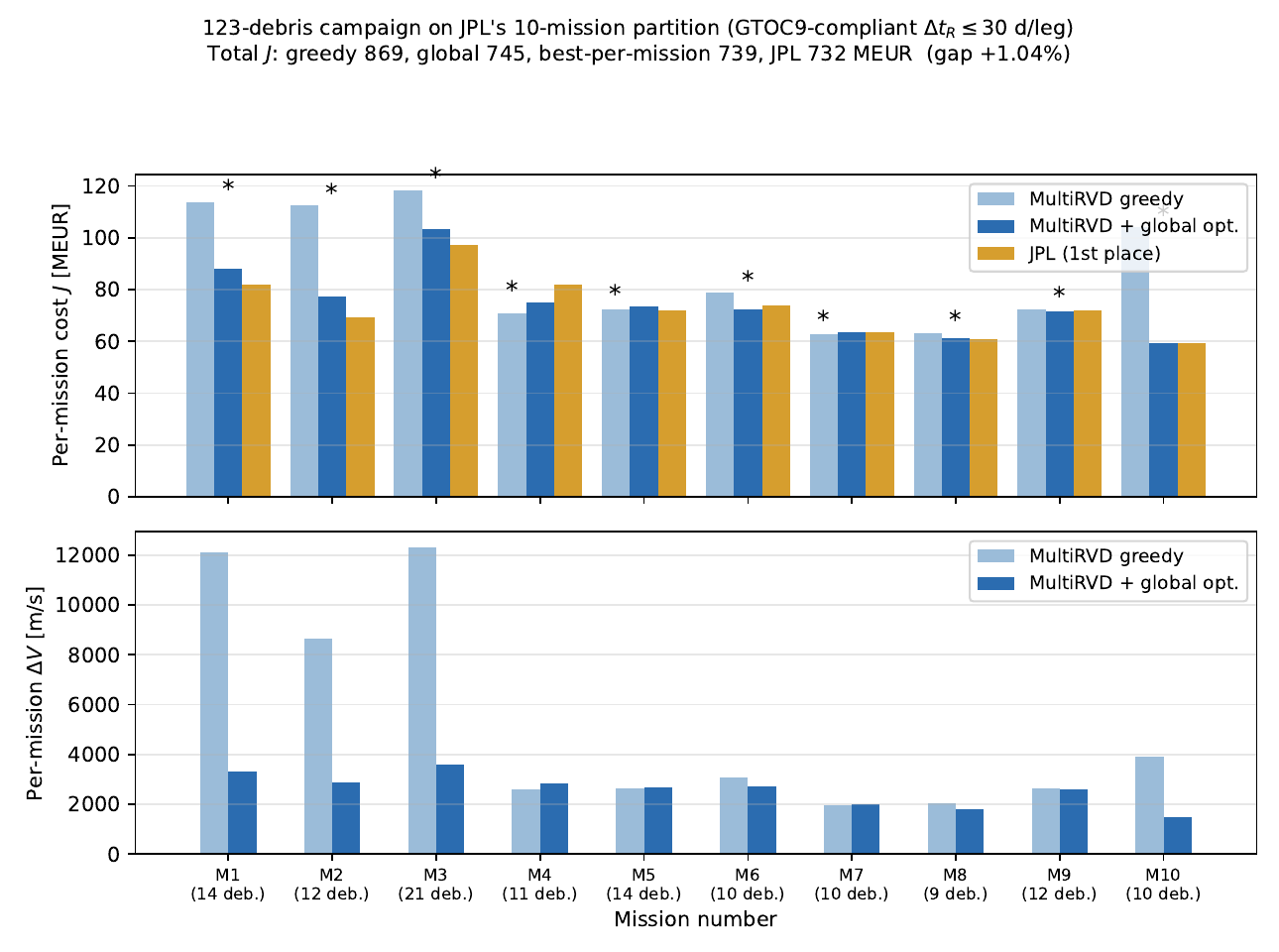}
\caption{Full 123-debris campaign benchmark context: per-mission cost
$J$ (top) and per-mission $\Delta V$ (bottom) for the present
framework's greedy and global-optimised variants under the
GTOC9-compliant $\Delta t_R\!\leq\!30$~d/leg constraint, shown alongside
the published JPL per-mission reference values for comparison.}
\label{fig:full_campaign}
\end{figure*}

For each of the 10 missions in the published JPL partition we extract
the corresponding debris subset (9--21 objects/mission, Table~1
of~\cite{b_jpl}), propagate every debris's RAAN from its catalogue
reference epoch to the reported JPL mission start epoch
via~\eqref{eq:raan_rate}, and evaluate the published visit sequence with
the greedy and globally-optimised pipelines under the GTOC9-compliant
$\Delta t_R\!\leq\!30$~d/leg constraint. Per-mission propellant
requirements are sized by inverse Tsiolkovsky and converted to mission
cost via $J = c_i + \alpha(m_0-m_{dry})^2$ with $c_i=55$~MEUR and
$\alpha=2\!\times\!10^{-6}$~MEUR/kg$^2$. Table~\ref{tab:full_campaign}
and Fig.~\ref{fig:full_campaign} report the per-mission results.

\begin{table*}[t]
\caption{Full 123-debris campaign on the published JPL mission partition
under the GTOC9-compliant $\Delta t_R\!\leq\!30$~d/leg constraint
(30-d transfer segment; Sec.~\ref{sec:problem}). Costs use the corrected
staged mass model that charges the first de-orbit package (30~kg) to launch
mass uniformly across the greedy, global, and best variants (per-mission
increments 0.15--0.58~MEUR), matching the journal companion's
convention~\cite{iskender_journal2026};
all $\Delta V$/cost entries are \emph{analytic-ledger} fidelity (two-impulse
Hohmann accounting), for which Sec.~\ref{sec:gmat} reports the measured
executable-decomposition overhead. Bold rows mark missions where the
analytical framework returns a cost lower than the published JPL reference
value; the ``Gap'' column reports the residual relative to that published
reference.}
\label{tab:full_campaign}
\centering
\footnotesize
\setlength{\tabcolsep}{4pt}
\begin{tabular}{@{}rrrrrrrrrr@{}}
\toprule
\textbf{Mis.} & \textbf{$n_{deb}$} & \multicolumn{2}{c}{\textbf{Greedy}} &
\multicolumn{2}{c}{\textbf{$+$Global opt.}} & \textbf{Best} &
\multicolumn{2}{c}{\textbf{JPL ref.~\cite{b_jpl}}} & \textbf{Gap} \\
\cmidrule(lr){3-4}\cmidrule(lr){5-6}\cmidrule(lr){8-9}
 & & $\Delta V$ & $J$ & $\Delta V$ & $J$ & $J$ & $m_0$ & $J$ & $\Delta J$ \\
 & & [m/s] & [MEUR] & [m/s] & [MEUR] & [MEUR] & [kg] & [MEUR] & [\%] \\
\midrule
1  & 14 & 12\,106 & 114.3 &  3\,321 &  88.5$^\dagger$ &  88.5 & 5\,665 & 81.9 & $+8.1$  \\
2  & 12 &  8\,643 & 112.9 &  2\,883 &  77.9$^\dagger$ &  77.9 & 4\,666 & 69.2 & $+12.5$ \\
3  & 21 & 12\,320 & 119.0 &  3\,610 & 103.8$^\dagger$ & 103.8 & 6\,590 & 97.1 & $+6.8$  \\
\textbf{4}  & 11 &  2\,597 &  71.0$^\ddagger$ &  2\,849 &  75.3 & \textbf{71.0} & 5\,679 & 82.1 & $\mathbf{-13.4}$ \\
5  & 14 &  2\,623 &  72.7$^\ddagger$ &  2\,690 &  73.6 &  72.7 & 4\,907 & 71.9 & $+1.1$  \\
\textbf{6}  & 10 &  3\,076 &  79.2 &  2\,721 &  72.7$^\dagger$ & \textbf{72.7} & 5\,063 & 73.8 & $\mathbf{-1.5}$ \\
\textbf{7}  & 10 &  1\,949 &  63.0$^\ddagger$ &  2\,013 &  63.7 & \textbf{63.0} & 4\,082 & 63.7 & $\mathbf{-1.1}$ \\
8  &  9 &  2\,051 &  63.5 &  1\,818 &  61.6$^\dagger$ &  61.6 & 3\,726 & 61.0 & $+1.0$  \\
9  & 12 &  2\,624 &  72.8 &  2\,581 &  72.1$^\dagger$ &  72.1 & 4\,897 & 71.8 & $+0.4$ \\
10 & 10 &  3\,906 & 104.3 &  1\,479 &  59.5$^\dagger$ &  59.5 & 3\,439 & 59.1 & $+0.5$  \\
\midrule
\textbf{Tot.} & \textbf{123} & \textbf{51\,893} & \textbf{872.7} &
\textbf{25\,965} & \textbf{748.7} & \textbf{742.65} & n/a &
\textbf{731.5} & $\mathbf{+1.52}$ \\
\bottomrule
\multicolumn{10}{l}{\scriptsize $^\dagger$ Best picks the global-opt result; $^\ddagger$ best picks the greedy result.} \\
\multicolumn{10}{l}{\scriptsize JPL paper reports 731.28~MEUR; 731.5 here is recovered from per-mission masses (0.03\% rounding). Campaign best $=742.65$~MEUR, $+1.55\%$ vs.\ 731.28.} \\
\multicolumn{10}{l}{\scriptsize Totals from unrounded per-mission values; columns may not sum exactly. The M10 campaign global-opt $\Delta V$ (1\,479~m/s) differs from the} \\
\multicolumn{10}{l}{\scriptsize \quad deep-dive figure (1\,484~m/s, Table~\ref{tab:global_opt}) \emph{not} by any stochastic seed (both fix DE seed~42 and are deterministic) but because the} \\
\multicolumn{10}{l}{\scriptsize \quad campaign uses a reduced optimiser config (maxiter~60, popsize~15, 2 CD passes) vs.\ the deep-dive (maxiter~100, popsize~20, 10 passes).} \\
\end{tabular}
\end{table*}

Three observations stand out. \emph{First}, the global time-budget
optimizer reduces total campaign $\Delta V$ from 51\,893~m/s (greedy) to
25\,965~m/s, a 50\% drop driven principally by Missions~1, 2, 3, and~10,
where greedy allocation funnels several legs into the costly
direct-transfer regime. \emph{Second}, taking the lower-$J$ variant per
mission yields 742.65~MEUR against the published JPL reference
value of 731.5~MEUR (JPL's submitted competition score of 731.28~MEUR
in the published paper), a residual of
$+1.52\%$ ($+1.55\%$ against the submitted 731.28~MEUR), demonstrating
that the analytical framework reproduces the published benchmark's
overall cost level to within about one and a half percent. This total
charges the first de-orbit package on every mission in the staged mass
sizing; the earlier 739.1~MEUR figure omitted that package and is
superseded. \emph{Third}, the per-mission breakdown shows
that for three missions (M4, M6, M7) the framework returns a cost
below the published reference value, and three more (M8, M9, M10)
lie within about one percent of the published reference; Mission~9,
marginally below the reference before the mass-sizing correction,
now sits just above it at $+0.4\%$ and no longer beats JPL, whereas
the M4/M6/M7 margins survive. The residual aggregate
gap is concentrated on Missions~1--3, the longest and most debris-dense
missions where the semi-analytic multi-impulse transfers and SNOPT
refinement used in the JPL solution offer fidelity that two-impulse
Hohmann transfers cannot capture. Taken together, the per-mission cost trends are broadly
consistent with the published reference, which is the validation
outcome being sought here.

\paragraph{Benchmark context against published GTOC9 reference solutions.}
Table~\ref{tab:benchmark_context} situates the result against published GTOC9
reference solutions, providing benchmark context for the analytical
framework's overall cost level. Because the present study adopts the
published JPL mission partition, the comparison should be read as a
fidelity check on the trajectory-optimization layer at a fixed
partition rather than as an independent end-to-end comparison of methods.
Note also that 731.28~MEUR is JPL's \emph{submitted} competition score;
their paper reports post-competition ten-mission improvements to
approximately 720 and 711~MEUR~\cite{b_jpl}, and the submitted value is
retained as the benchmark anchor throughout this paper.

\begin{table*}[t]
\caption{Final 10-mission campaign cost in the context of published
GTOC9 reference solutions. The ``Rel.\ JPL ref.'' column reports the
residual relative to the published JPL reference campaign cost
(adopted as the trusted benchmark anchor in this study). Both
present-framework rows adopt JPL's partition and identical per-leg analytic
transfers, differing only in time allocation: greedy per-leg vs.\ global
time-budget optimization. Entries other
than the present framework's are quoted directly from the corresponding
publications.}
\label{tab:benchmark_context}
\centering
\footnotesize
\begin{tabular}{@{}lrlrr@{}}
\toprule
\textbf{Study / variant} & \textbf{Missions} & \textbf{Approach} &
\textbf{$J$ [MEUR]} & \textbf{Rel.\ JPL ref.} \\
\midrule
Published JPL ref.~\cite{b_jpl} & 10 & Semi-analytic 6-impulse database $+$ B\&B/ACO $+$ SNOPT & 731.28 & n/a     \\
\textbf{Present framework}      & \textbf{10} &
\textbf{Analytic Hohmann, min$(A,B,C^+)$, global time opt.} &
\textbf{742.65} & $\mathbf{+1.55\%}$ \\
Published NUDT~\cite{b_nudt}    & 12 & 3-level decomp.\ $+$ DCB-ACO    & 786.21 & $+7.5\%$ \\
Published Tsinghua~\cite{b_tsinghua} & 12 & PSO impulsive $+$ beam search & 829.58 & $+13.4\%$ \\
Present framework (greedy only) & 10 & Analytic Hohmann, min$(A,B,C^+)$, greedy time alloc.   & 872.70 & $+19.3\%$ \\
Published Strathclyde~\cite{b_strathclyde} & 14 & Beam search $+$ Time Shuffler (MACS) & 918.98 & $+25.7\%$ \\
Published DLR~\cite{b_dlr}      & 14 & Time-dependent TSP $+$ beam search & 949.85 & $+29.9\%$ \\
\bottomrule
\end{tabular}
\end{table*}

The 742.65~MEUR total remains within about one and a half percent of the
published JPL reference campaign cost while using only analytical two-impulse
transfers and a $\sim$27-minute single-core pipeline; this is the
methodological observation of interest, namely that a lightweight
analytical model exploiting $J_2$ as a resource captures the bulk of
the cost behaviour observed in the published reference solution that
relies on a $\sim$290-million-row precomputed semi-analytic six-impulse
transfer database and SNOPT multi-impulse refinement~\cite{b_snopt}. The comparison should not be interpreted as a ranking of
methods, since the partition decision (a substantial part of the JPL
contribution) is reused verbatim.

\subsection{Side Validations: SSO Inspection and Commercial Tug}
\label{subsec:other_cases}

Two side scenarios validate transferability outside GTOC9. Both use
\emph{synthetic} target catalogues constructed to span the stated RAAN
bands, not flight datasets, so they probe the method's behaviour across
mission archetypes rather than reproduce a specific mission.

\paragraph{SSO inspection.}
A 3\,000~kg servicer with $I_{sp}=320$~s visits 10 defunct satellites in a
$\sim$27$^\circ$ RAAN band. The RAAN-sorted and greedy NN methods produce
identical sequences; the entire 32.9\% $\Delta V$ saving
(4\,146 to 2\,782~m/s) comes from drift-aware \emph{timing}, not
sequencing. The dominant cost lever in this regime is Method~B
(wait-for-drift) on a tightly-clustered catalogue, with C+ providing no
incremental benefit because RAAN gaps are too small.

\paragraph{Commercial space tug.}
A reusable tug ($m_{wet}=5\,500$~kg, $I_{sp}=330$~s) delivers 6 client
satellites from a 500~km parking orbit to SSO targets spanning
$10^\circ$--$35^\circ$ in RAAN. The round-trip architecture exposes the
tug to long inter-delivery drift periods (970~days total). Greedy NN with
drift exploitation reduces $\Delta V$ by 81.4\%
(10\,334 to 1\,921~m/s) versus the direct-transfer baseline; against the
operationally relevant Method-B baseline (passive drift, 2\,332~m/s),
active C+ shaping adds a further 17.5\% saving.

\subsection{Numerical Validation}
\label{sec:gmat}

The model-fidelity claims above are validated in two stages. First, the
in-house FPROX numerical propagator is benchmarked against NASA's General
Mission Analysis Tool (GMAT) R2025a~\cite{b_gmat} to establish its
adequacy as a $J_2$-fidelity reference; then FPROX is used to verify
the framework's per-leg drift-orbit manoeuvre plan across every leg of
the JPL mission partition. The validation harness is documented in
the project repository under \texttt{validation/GMAT/comparison/}.

\begin{figure}[ht]
\centering
\includegraphics[width=\columnwidth]{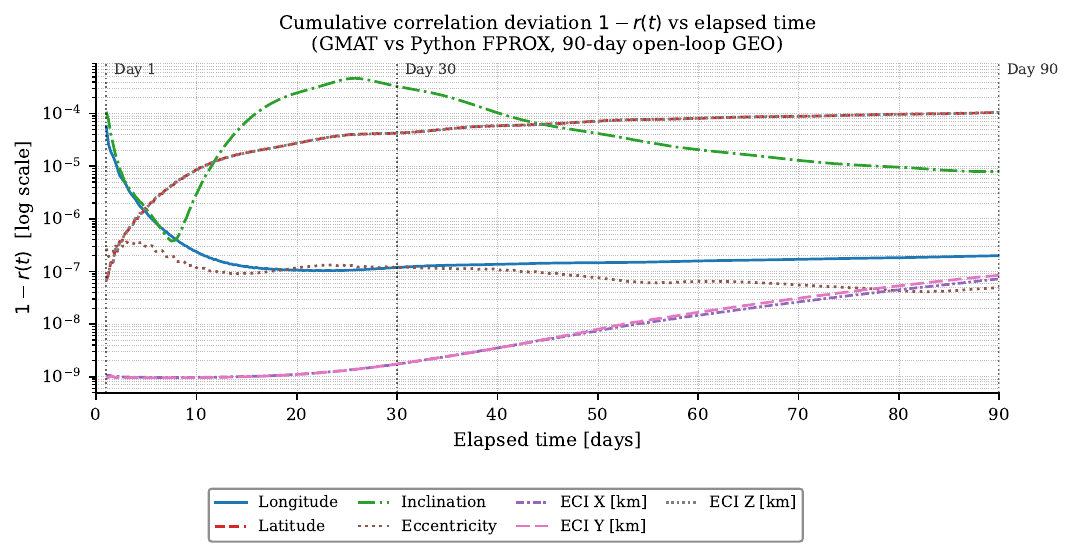}
\caption{Cumulative correlation deviation $1{-}r(t)$ vs.\ elapsed time for
  the seven state channels (3 position, 3 velocity, longitude), FPROX
  vs.\ GMAT. Vertical dotted lines mark the Day~1, Day~30, and Day~90
  checkpoints. All channels stay below $10^{-4}$ throughout the 90-day
  propagation.}
\label{fig:fprox_gmat_corr}
\end{figure}

\paragraph{FPROX$\leftrightarrow$GMAT propagator benchmark.}
FPROX (DOP853, adaptive) was benchmarked against GMAT across five
90-day open-loop scenarios (Table~\ref{tab:fprox_gmat}) spanning two
epochs, three eccentricities, two inclinations, and a three-burn
impulsive case. Initial conditions are
taken directly from the GMAT Cartesian state at $t{=}0$ so that
residuals reflect only force-model and integrator divergence; the two
propagators are compared on a uniform 6-hour cubic-spline grid
(361 points per scenario). The 3D position RMS stays below 1~km for the
four ballistic cases and 1.6~km for the impulsive-burn case where
along-track drift after the cross-track burn dominates. Per-channel
agreement is further measured via the Pearson correlation coefficient~$r$;
the cumulative deviation $1{-}r(t)$ (Fig.~\ref{fig:fprox_gmat_corr})
stays below $10^{-4}$ in every channel through Day~90, confirming that
FPROX is a quantitatively faithful surrogate for GMAT under the $J_2$
force model on time scales comparable to a GTOC9 leg
(30~d) or mission (273~d).

\begin{table*}[ht]
\centering
\caption{FPROX-Python (DOP853) vs.\ GMAT R2025a benchmark across five
  90-day open-loop scenarios. IC mismatch is identically zero; residuals
  reflect independent force-model and integrator implementations.
  Comparison grid: 6-hour cubic spline (361 points).}
\label{tab:fprox_gmat}
\footnotesize
\begin{tabular}{clcccc}
\toprule
Sc & Configuration & \makecell{3D RMS\\{[km]}} & \makecell{3D Max\\{[km]}}
   & \makecell{Inc RMS\\{[deg]}} & \makecell{worst $1{-}r$} \\
\midrule
1 & JAN2024 ballistic, $e{=}10^{-4}$, $i{=}0.05^\circ$
  & 0.898 & 2.07 & $8.0{\times}10^{-4}$ & $1.1{\times}10^{-4}$ \\
2 & JUN2019 ballistic, $e{=}10^{-4}$, $i{=}0.05^\circ$
  & 0.680 & 1.53 & $7.0{\times}10^{-4}$ & $4.5{\times}10^{-5}$ \\
3 & JUN2019 ballistic, $e{=}3{\times}10^{-4}$, $i{=}0.05^\circ$
  & 0.680 & 1.53 & $7.0{\times}10^{-4}$ & $4.5{\times}10^{-5}$ \\
4 & JUN2019 ballistic, $e{=}3{\times}10^{-4}$, $i{=}5^\circ$
  & 0.904 & 1.78 & $5.6{\times}10^{-4}$ & $1.3{\times}10^{-10}$ \\
5 & JUN2019 + 3 impulsive burns, $i{=}5^\circ$
  & 1.640 & 2.91 & $5.4{\times}10^{-4}$ & $6.8{\times}10^{-10}$ \\
\bottomrule
\end{tabular}
\end{table*}

\paragraph{Spot-check: secular RAAN rate.}
For the GTOC9 setting specifically, GMAT was run as a direct osculating
reference for one representative debris orbit (\#5: $a=7228.5$~km,
$e=1.6\!\times\!10^{-3}$, $i=99.84^\circ$), propagated 30~days under a
$J_2$-only force model. The secular RAAN rate derived from the GMAT
Brouwer-mean time series is $\dot{\Omega}_{\text{GMAT}} = 1.140^\circ$/day
versus the closed-form value $\dot{\Omega}_{\text{analytic}} =
1.099^\circ$/day from~\eqref{eq:raan_rate}, a $+3.8\%$ offset. The gap
is shown empirically to be independent of higher zonals ($J_3$--$J_8$
contribute $<0.2\%$), short-period oscillation (Brouwer-mean and
osculating drifts agree to 5~m$^\circ$), and constants (JGM-2 matches
GTOC9's $\mu$, $r_{eq}$, $J_2$ to eight digits). Having ruled those out, the residual is attributed to a reference-frame effect
between the epoch at which the debris elements are defined and the epoch at
which GMAT evaluates the geopotential. Over the multi-decade gap between the
J2000 geopotential frame (in which the catalogue elements are given) and the
propagation epoch (about whose true equator GMAT evaluates the field), IAU
precession tilts Earth's symmetry axis by a fraction of a degree; because
$\dot\Omega\propto\cos i$ and $|\tan i|$ is large in the near-polar
Sun-synchronous regime, a first-order estimate of that frame tilt accounts for
the observed $\sim$3.8\% change in $|\dot\Omega|$. The offset is therefore a
first-order reference-frame effect, not a deficiency of the analytic $J_2$
model and not an additional $\Delta V$ cost: it is a small, bounded difference
in the nodal-regression rate that does not enter the transfer-cost ledger, and
the realised-cost driver is instead the measured four-burn executable overhead
quantified below.

\paragraph{Campaign-wide per-leg plane-acquisition verification (FPROX).}
FPROX is the framework's \emph{own} self-contained numerical-$J_2$
propagator, here cross-checked against a single-leg GMAT run rather than
treated as a GMAT surrogate; it is used to verify the framework's
3-knob C$^{+}$ manoeuvre plan on every leg of the JPL mission
partition (113 legs across 10 missions). For each leg the framework's
analytically-prescribed four impulsive burns (entry Hohmann pair with
plane change at the orbital line of nodes, drift coast, exit Hohmann
pair) are applied verbatim in $J_2$-only numerical propagation; the
post-manoeuvre chaser state is then compared in Brouwer-mean elements
against the target debris. The verification closes the \emph{orbit plane
and altitude} $(a,i,\Omega)$ only: the GTOC9 catalogue discards mean
anomaly at load, so the in-track phase/anomaly is never scored, and a
terminal along-track phasing manoeuvre (standard, low-$\Delta V$, and
time-consuming) is assumed but not simulated here. Across the 89 legs
where the C$^{+}$ planner returns a feasible drift orbit (24~legs fall
back to method~B and are excluded), the realised RAAN-miss distribution
has median $0.136^\circ$ and 83\% of legs acquire the target plane to
better than 1$^\circ$; relative to an ideal-impulse upper bound on the
same propagator, the median ratio is~1.15$\times$, i.e.\ the analytic
plan is within 15\% of the ideal plane-acquisition floor on the typical
leg: the plane and altitude are acquired, with along-track phasing
deferred to the unsimulated terminal manoeuvre.

\paragraph{Executable-decomposition overhead.}
The headline $\Delta V$ and cost of
Secs.~\ref{subsec:m10}--\ref{subsec:full_campaign} are stated at
\emph{analytic-ledger} fidelity, the two-/four-impulse Hohmann accounting.
Decomposing each leg into its physically-executable four-burn plan and
re-summing over the 89 verified legs raises the total from
33\,626 to 36\,222~m/s, a \emph{measured} $+7.7\%$ campaign-wide executable
overhead ($+8.9\%$ on the median leg) above that ledger. This is a
validation result, not an assumed penalty.

\paragraph{Wide-gap tail.}
The 15 wide-gap legs that miss by more than 1$^\circ$ are 17\% of the 89
verified legs but carry 23.6\% of the verified-leg $\Delta V$; they miss by
a median $5.7^\circ$ (max $10.55^\circ$) in RAAN and up to $1.78^\circ$ in
inclination. Closing them requires a dedicated plane-targeting burn whose
cost is \emph{not} quantified here and reaches of order $1$~km/s on the
worst leg, the regime where the analytic small-angle linearisation breaks
down and where multi-impulse numerical refinement, as in the published JPL
solution, retains a clear advantage.

\paragraph{Implications.}
Two fidelity gaps separate the analytic-ledger headline from a literal
high-fidelity flight, and round-2 replaces the earlier loose
extrapolation with reproducible, artifact-backed bounds. First, the
physically-executable four-burn decomposition adds a \emph{measured}
$+7.7\%$ to the campaign $\Delta V$ (above); applied to the mass model this
scales the analytic 742.65~MEUR ledger only modestly and reproducibly, and
must not be read as ``within the spread of published solutions''; it is a
sensitivity computed inside the present pipeline, not an ESA-verified score.
Second, the larger uncertainty is the wide-gap tail: on the 15 outlier legs
a dedicated plane-targeting burn (unquantified here, of order $1$~km/s on
the worst leg) would be required, and on the debris-dense Missions~1--3 the
associated propellant growth can push a literal execution against the
5\,000-kg GTOC9 tank capacity, at which point the plan must be
\emph{re-optimised} (per-leg time budgets, and potentially the partition)
rather than merely re-costed. The $+3.8\%$ nodal-rate offset is, as discussed above, a first-order
reference-frame effect between the catalogue epoch and the propagation epoch,
not a deficiency of the analytic model and not an additional $\Delta V$ cost; it
is a small, bounded rate error rather than a missing force, and the realised
execution cost is instead driven by the measured $+7.7\%$ four-burn overhead. All of these are extrapolations within the present
pipeline, not end-to-end high-fidelity propagations, and they must
\emph{not} be read against the ESA-verified published GTOC9 team costs.

\section{Discussion}
\label{sec:discussion}

\paragraph{Model fidelity gap.}
Under identical constraints (the published JPL Mission~10 reference
sequence, $\leq$16~d/leg transfer time), the proposed hybrid achieves
1\,669~m/s, approximately 17\% above the published JPL reference value
of 1\,423~m/s. This residual arises from three sources: (i)~two-impulse analytic
Hohmann transfers (four impulses on drift legs) vs.\ the semi-analytic
six-impulse database transfers of the published JPL solution,
SNOPT-refined with up to ten impulses per revolution~\cite{b_snopt};
(ii)~discrete $(e_d,i_d)$ grid search with
$a_d$ by bisection vs.\ continuous nonlinear optimization; and
(iii)~the circular-orbit assumption ignoring debris eccentricities up
to $e\!=\!0.02$. When the global time-budget optimizer is applied
under the GTOC9-compliant $\leq$30~d/leg cap, the residual narrows to
4.3\% (1\,484~m/s); the relaxed-time variant (1\,434~m/s, $+0.8\%$) is
not a GTOC9-legal solution and is reported solely for sensitivity
analysis. Either way, the primary source of the 17\% residual under
greedy allocation is suboptimal per-leg time budgeting, not
transfer-model fidelity per~se, an engineering finding that motivates
the global time-budget allocator developed in
Sec.~\ref{sec:methods}.

\paragraph{Where the analytical model matches the published reference well.}
The three missions on which the framework returns a cost below
the published JPL reference value (M4, M6, M7) share a common
feature: their per-leg RAAN gaps are small enough that drift exploitation
closes them passively, so the analytic Hohmann ledger captures essentially
the same physics as the semi-analytic multi-impulse transfers used in the
published JPL reference. The residual aggregate gap on M1--M3 is driven
by one to three outlier legs per mission with $\Delta\Omega\!>\!2^\circ$
that the 30-day per-leg cap cannot close passively; the regime where
multi-impulse SNOPT refinement, as deployed in the published JPL
solution, retains a clear modelling advantage and where future hybrid
analytic--numerical workflows should focus. Interpreted as an engineering
observation, this maps cleanly to where multi-impulse refinement matters
and where it does not.

\paragraph{Computational efficiency.}
Method~A evaluates in microseconds; Method~B requires $\sim$0.1~ms per
day of wait time; Method~C+ performs 2D grid search with bisection at
$\sim$50--500~ms per leg. The full 10-mission greedy campaign completes
in 87~s on a single CPU core; adding the global time-budget optimizer
per mission raises the wall-clock to approximately 27~minutes for the
full 123-debris campaign. For context, the published JPL solution reports
precomputation of a $\sim$290-million-row semi-analytic six-impulse
transfer database; the analytical
formulation here therefore operates at a fundamentally different
computational scale. These evaluation speeds support interactive
trade-space exploration and in-flight replanning, capabilities that are
impractical with precomputed transfer databases and that the
analytical pipeline is naturally suited to deliver.

\paragraph{Limitations.}
The impulsive maneuver model precludes application to low-thrust electric
propulsion. Atmospheric drag, solar radiation pressure, and third-body
perturbations are neglected (secondary effects at GTOC9 altitudes
620--920~km). Target tumbling, capture dynamics, and rendezvous proximity
operations are not modelled. Navigation and execution errors are also
neglected; incorporating $\Delta V$ margins and stochastic analysis would
provide more conservative estimates. The circular-orbit assumption
introduces small errors for the GTOC9 catalogue ($e\!\leq\!0.02$). The
journal companion~\cite{iskender_journal2026} carries the extended
per-leg numerical verification that quantifies how these idealisations translate into
realised rendezvous miss residuals and execution-cost overhead. The
complementary close-proximity rendezvous-and-docking problem (the
terminal phase that converts the residual miss into a docked
configuration) has been treated thoroughly and experimentally in the
authors' prior work~\cite{iskender2019dual,burak2020phd} and is outside
the multi-rendezvous scope of the present paper.

\section{Conclusion}
\label{sec:conclusion}

This paper presented a drift-aware multi-rendezvous trajectory
optimization framework that systematically exploits $J_2$-induced RAAN
precession as a mission-design resource, validated using GTOC9 as a
controlled benchmark environment and the published JPL solution as a
trusted reference baseline. Three principal methodological results are
reported. The three-parameter drift orbit (C+) reduces per-leg transfer
cost by up to ${\sim}40\%$ on favorable legs versus altitude-only
manipulation for SSO orbits, with
inclination as the dominant lever at near-polar inclinations. The
sequence--timeline coupling (per-leg drift time shifting
downstream RAAN geometry through accumulated elapsed time) is made
explicit and resolved by a two-phase global time-budget optimizer. The
full 123-debris GTOC9 benchmark on the published JPL mission
partition under the GTOC9-compliant $\Delta t_R\!\leq\!30$~d/leg
constraint yields 742.65~MEUR against the published JPL
reference value of 731.28~MEUR ($+1.55\%$; JPL's submitted competition
score, retained as the anchor even though their paper reports
post-competition improvements near 720 and 711~MEUR; the cost charges the
first de-orbit package on every mission), demonstrating that a
lightweight analytical pipeline reproduces the cost level of the
published reference to within about one and a half percent while running
in $\sim$27~minutes on a single CPU core. On three of the ten missions
(M4, M6, M7) the framework's per-mission cost lies below the published
reference value, consistent with the structural cost profile of the
published JPL benchmark on those missions. An independent
sequence-recovery validation reproduces a Mission~10 removal sequence
consistent with the published JPL reference from the debris catalogue
alone, and the analytic plan is numerically verified leg-by-leg with the
framework's own numerical-$J_2$ propagator (cross-checked against a
single-leg GMAT run): 83\% of the 89 verified legs acquire the target
orbit plane to better than 1$^\circ$ in RAAN (median 0.136$^\circ$),
within 15\% of the ideal plane-acquisition floor, with a terminal
along-track phasing manoeuvre assumed but not simulated. A 17\% tail of
wide-gap legs still needs a dedicated plane-targeting burn, and the
physically-executable four-burn decomposition adds a measured $+7.7\%$ to
the campaign $\Delta V$ above the analytic ledger, the principal realised-cost
driver; the documented $+3.8\%$ nodal-rate offset is a small, bounded
first-order reference-frame effect, not a model deficiency and not an added
$\Delta V$ cost.

Future work addresses four directions: (i)~joint partition-and-sequence
discovery via the Beam P-ACO backbone of Sec.~\ref{subsec:setcover}, in
which the mission partition itself is constructed alongside the
per-mission sequences rather than adopted as a benchmark reference;
(ii)~extension to low-thrust electric propulsion, where the
continuous-thrust arc replaces impulsive maneuvers and drift
exploitation requires reformulation as an optimal-control problem;
(iii)~multi-agent fleet coordination, where multiple spacecraft
partition the catalogue and must avoid temporal conflicts; and
(iv)~hybrid-fidelity refinement, where analytical drift-orbit solutions
seed full-dynamics numerical optimization in GMAT or equivalent,
potentially closing the residual relative to the published JPL
reference.

\section*{Acknowledgments}

The authors acknowledge ESA for organizing the GTOC9 Kessler Run
competition and making the problem data publicly available. The first
author acknowledges Nanyang Technological University for research
support. The authors thank Final Proximity Space Systems for their
support.

\balance
\bibliographystyle{unsrt}
\bibliography{references}

\end{document}